\documentclass[amsmath,amssymb,aps,prl,twocolumn,superscriptaddress,nofootinbib]{revtex4-1}
\usepackage{graphicx,color,tikz,pgf,csquotes,hyperref,braket,bbm,physics,tensor,upgreek}
\usepackage{enumitem}

\newcommand{\one}{\mathbbm 1}
\newcommand{\R}{\mathbb R}
\newcommand{\C}{\mathbb C}
\newcommand{\e}{\textrm{e}}

\newcommand{\SL}{\text{SL$(2,\C)$}}

\newcommand{\qmarks}[1]{``#1''}

\newcommand{\GFT}{\mathrm{GFT}}
\newcommand{\GR}{\mathrm{GR}}

\newcommand{\Pl}{\mathrm{Pl}}
\newcommand{\vmf}{\boldsymbol{\psi}}

\newcommand{\mf}{\phi} 
\newcommand{\mm}{\pi_\phi} 
\newcommand{\pmm}{p_\phi} 
\newcommand{\rf}{\chi} 

\newcommand{\pip}{\pi_0^+}

\newcommand{\tpip}{\tilde{\pi}_0^+}
\newcommand{\tpim}{\tilde{\pi}_0^-}
\newcommand{\slrcw}{\tilde{\sigma}} 
\newcommand{\tlrcw}{\tilde{\tau}} 

\begin{document}

\title{Scalar Cosmological Perturbations from Quantum Gravitational Entanglement}

\author{Alexander F. Jercher}
\email{alexander.jercher@uni-jena.de}
\affiliation{
Arnold Sommerfeld Center for Theoretical Physics,
Ludwig-Maximilians-Universit\"at M\"unchen,\\
Theresienstr.~37, 
80333 M\"unchen, Germany, EU
}
\affiliation{
Munich Center for Quantum Science and Technology (MCQST),\\ Schellingstr.~4, 
80799 M\"unchen, Germany, EU
}
\affiliation{Theoretisch-Physikalisches Institut, Friedrich-Schiller-Universit\"{a}t Jena,\\ Max-Wien-Platz~1, 07743 Jena, Germany, EU
}
\author{Luca Marchetti}
\email{luca.marchetti@unb.ca}
\affiliation{Department of Mathematics and Statistics, University of New Brunswick,\\ Fredericton, NB, Canada E3B 5A3
}
\author{Andreas G. A. Pithis}
\email{andreas.pithis@uni-jena.de}
\affiliation{
Arnold Sommerfeld Center for Theoretical Physics,
Ludwig-Maximilians-Universit\"at M\"unchen,\\
Theresienstr.~37, 
80333 M\"unchen, Germany, EU
}
\affiliation{
Munich Center for Quantum Science and Technology (MCQST),\\ Schellingstr.~4, 
80799 M\"unchen, Germany, EU
}
\affiliation{Theoretisch-Physikalisches Institut, Friedrich-Schiller-Universit\"{a}t Jena,\\ Max-Wien-Platz~1, 07743 Jena, Germany, EU
}
\affiliation{Center for Advanced Studies (CAS), Ludwig-Maximilians-Universit\"at München,\\ Seestr.~13, 80802 München, Germany, EU
}

\begin{abstract}
A major challenge at the interface of quantum gravity and cosmology is to explain the emergence of the large-scale structure of the Universe from Planck scale physics. In this letter, we extract the dynamics of scalar isotropic cosmological perturbations from full quantum gravity, as described by the causally complete Barrett-Crane group field theory model. From the perspective of the underlying quantum gravity theory, cosmological perturbations are represented as nearest-neighbor two-body entanglement of group field theory quanta. Their effective dynamics is obtained via mean-field methods and described relationally with respect to a causally coupled physical Lorentz frame. We quantitatively study these effective dynamical equations and show that at low energies they are perfectly consistent with those of General Relativity, while for trans-Planckian scales quantum effects become important. These results therefore not only provide crucial insights into the potentially purely quantum gravitational nature of cosmological perturbations, but also offer rich phenomenological implications for the physics of the early Universe.
\end{abstract}

\maketitle

The extraction of cosmology from full quantum gravity (QG) constitutes a fundamental objective shared across all quantum gravity approaches~\cite{Blumenhagen:2013fgp,Percacci,Reuter,Bonanno:2020bil,Ambjorn:2012jv,Hamber2009,Surya:2019ndm,Rovelli:2004tv,Rovelli:2011eq,Perez:2012wv,Asante:2022dnj,Ashtekar:2004eh,Thiemann:2007pyv,Freidel:2005qe,Oriti:2011jm,Carrozza:2013oiy,Gurau:2016cjo,GurauBook}. This endeavor holds the promise of providing a UV-completion of cosmological models, potentially resolving longstanding open questions of the $\Lambda$CDM model~\cite{Planck:2018vyg}, in particular concerning the origin of primordial perturbations, and subjecting quantum gravity theories to the testing ground of cosmological observations. Achieving this goal, however, presents a formidable challenge, particularly in approaches with fundamental microscopic degrees of freedom that differ significantly from the continuum gravitational field\footnote{See Ref.~\cite{Oriti:2013jga} for a discussion on the theoretical evidence supporting this hypothesis.} at the heart of standard cosmology. This challenge involves the definition of a robust coarse-graining procedure to identify macroscopic geometric and matter degrees of freedom, and to define time evolution and spatiotemporal localization in the absence of a conventional spacetime manifold.

Group field theories (GFTs)~\cite{Freidel:2005qe,Oriti:2006se,Oriti:2011jm} provide a rich and versatile framework to tackle both of these challenges. They are non-local quantum field theories of spacetime, the fundamental excitations of which correspond to discrete quantum geometric building blocks. More precisely, GFTs are closely connected to canonical loop quantum gravity (LQG)~\cite{Oriti:2013aqa,Ashtekar:2004eh} and provide a completion of the quantum dynamics of spin foam
models~\cite{Rovelli:2011eq,Perez:2012wv,Engle:2023qsu}. They also furnish a reformulation of specific simplicial lattice gravity path integrals~\cite{Bonzom:2009hw,Baratin:2010wi,Baratin:2011hp,Baratin:2011tx} and via their connection to tensor models~\cite{Gurau:2011xp,GurauBook,Gurau:2016cjo,Jercher:2022mky} further relate to the dynamical triangulations approach~\cite{Ambjorn:2012jv,Loll:2019rdj,Loll:1998aj}.

Spatially homogeneous and flat cosmological dynamics have been successfully recovered from the mean-field dynamics of GFT condensate states~\cite{Gielen:2013kla,Gielen:2013naa,Oriti:2016qtz,Gielen:2016dss,Pithis:2019tvp,Marchetti:2020umh,Jercher:2021bie}, representing the simplest form of coarse-graining of the underlying quantum gravitational dynamics. The existence of such states has recently been supported by a Landau-Ginzburg mean-field analysis of Lorentzian GFTs~\cite{Marchetti:2022igl,Marchetti:2022nrf,Dekhil:2024ssa,Dekhil:2024djp}. Moreover, a special class of condensate states, called coherent peaked states (CPSs)~\cite{Marchetti:2020qsq,Marchetti:2020umh,Marchetti:2021gcv}, provides a solution to the aforementioned localization problem by allowing the construction of effective relational observables.

Deriving the dynamics of cosmological perturbations from GFTs has been commenced in~\cite{Gielen:2017eco,Gielen:2018xph,Gerhardt:2018byq,Gielen:2022iuu} and in particular pushed forward in~\cite{Marchetti:2021gcv}, where the classical perturbation equations of General Relativity (GR) have been recovered in the super-horizon limit of large perturbation wavelengths. However, deviations from classical results occur there for non-negligible wave vectors, which we here argue to arise from an insufficient coupling between the physical reference frame and the underlying causal structure.

In this letter, we advance this program by deriving cosmological perturbation equations from GFT quantum gravity 
that agree with classical results at sub-Planckian scales and are subject to quantum corrections in the trans-Planckian regime. Within our framework,  inhomogeneities of cosmological observables emerge from entanglement between the quantum geometric degrees of freedom of the model. Thus, we also provide a concrete realization of the general expectation that non-trivial geometries are associated with quantum gravitational entanglement~\cite{Giddings:2005id,Ryu:2006bv,VanRaamsdonk:2010pw,Bianchi:2012ev,Maldacena:2013xja,Giddings:2015lla,Cao:2016mst,Donnelly:2016auv,swingle2018spacetime,colafrancheschi2022emergent,Bianchi:2023avf}. This is achieved as follows:
\begin{enumerate}[leftmargin=*]
    \item We utilize the rich causal structure of the completion~\cite{Jercher:2021bie,Jercher:2022mky} of the Barrett-Crane (BC) GFT model~\cite{Barrett:1999qw,Perez:2000ep,Perez:2000ec,Oriti:2000hh}, which provides a GFT quantization of first-order Palatini gravity, to causally couple a physical Lorentzian reference frame to the GFT model.
    \item We represent slightly inhomogeneous cosmological geometries via \emph{entangled states} $\ket{\Delta;x^0,\vb*{x}}$. We study their perturbative dynamics using mean-field methods \cite{Gielen:2013kla,Gielen:2013naa,Oriti:2016qtz,Gielen:2016dss,Pithis:2019tvp,Marchetti:2020umh,Jercher:2021bie,Marchetti:2021gcv}.
    \item We obtain effective relational cosmological observables by computing expectation values of appropriate operators on the states $\ket{\Delta;x^0,\vb*{x}}$. We study in detail the dynamics of cosmological inhomogeneities emerging from the above entanglement, determining their classical limit and characterizing quantum effects.
\end{enumerate}
We will discuss these steps in detail, one at a time, in each of the next three sections.

\section{The BC model with Lorentzian Frame}

We consider a minimal extension of the BC GFT model characterized by a spacelike ($+$) and timelike ($-$) sector~\cite{Jercher:2022mky}. The model is described by two fields $\varphi_\pm\equiv \varphi(g_v,X_\pm,\vmf)$.
Closure and simplicity constraints~\cite{Jercher:2022mky} are imposed on the $\varphi_\pm$, facilitating their excitations to be interpreted as quantum spacelike and timelike tetrahedra, respectively. Here, $g_v = (g_1,g_2,g_3,g_4)\in\SL^4$ and $X_\pm$ is a timelike ($+$), resp.~spacelike ($-$) normal vector to the tetrahedra~\cite{Jercher:2022mky}. Finally, $\vmf = (\chi^\mu,\mf)\in\R^5$ represents the values of $5$ minimally coupled massless free (MCMF) scalar fields discretized on dual vertices\footnote{The $\vmf$ are coupled so that the GFT Feynman amplitudes correspond to simplicial gravity path integrals with MCMF scalar fields~\cite{Oriti:2016qtz,Li:2017uao,Gielen:2018fqv}.}, which constitute the matter content of the system.

The dynamics of the model are specified by an action $S[\bar{\varphi},\varphi]=K[\bar{\varphi},\varphi]+V[\bar{\varphi},\varphi]$, where $K$ and $V$ represent kinetic and interaction terms. In particular, $K=K_++K_-$, with 
\begin{equation}
    K_\pm=\bar{\varphi}_{\pm}\cdot \mathcal{K}^{(X)}_\pm\cdot\varphi_\pm\,,
\end{equation}
where $\cdot$ represents an integration over the full GFT field domain. The bi-local kinetic kernels
\begin{equation}
    \mathcal{K}^{(X)}_\pm=\delta(X^{(v)}_\pm-X^{(w)}_\pm)\mathcal{K}_{\pm}(g_v,g_w;\Delta_{vw}^2\vmf)\,,
\end{equation}
where $\Delta^2_{vw}\vmf\equiv \left((\rf^\mu_v-\rf^\mu_w)^2,(\phi_v-\phi_w)^2\right)$, encode information about the propagation of geometry and matter data between neighbouring $4$-simplices (denoted $v$ and $w$) and clearly show that the $X_\alpha$ are non-dynamical. For further details, see~\cite{Jercher:2023nxa}.

The interaction $V$ is local in $\vmf$, non-local in the group theoretic variables and represents any possible gluing of five tetrahedra with arbitrary signature to form a $4$-simplex~\cite{Jercher:2022mky, Jercher:2023nxa}. 

The Fock space $\mathcal{F}$ of the model can be constructed by tensoring the Fock spaces associated to the two different sectors, i.e. $\mathcal{F}\equiv \mathcal{F}_+\otimes\mathcal{F}_-$, with
\begin{equation}
    \mathcal{F}_\pm \equiv \bigoplus_{N = 0}^\infty \mathrm{sym}\left(\mathcal{H}_\pm^{(1)}\otimes...\otimes \mathcal{H}_\pm^{(N)}\right),
\end{equation}
constructed out of the one-particle Hilbert spaces $\mathcal{H}_\pm$, see~\cite{Jercher:2023nxa} for further details. The $\mathcal{F}_\pm$ can be  generated by repeated action of the creation operators $\hat{\varphi}^\dagger_\pm\equiv \hat{\varphi}^\dagger_\pm \otimes \one_\mp$ on the Fock vacua $\ket{\emptyset}_\pm$ annihilated by $\hat{\varphi}_\pm$. Creation and annihilation operators satisfy
\begin{equation}
    \comm{\hat{\varphi}_{\pm}(\vmf)}{\hat{\varphi}_{\pm}^{\dagger}(\vmf')} = \delta^{(n)}(\vmf-\vmf')\one_{\pm}\,,
\end{equation}
where $\one_{\pm}$ represents the identity on $\mathcal{F}_\pm$ satisfying closure and simplicity constraints.

The four scalar fields $\rf^\mu$ with $\mu\in\{0,...,3\}$ will be used to construct a physical Lorentzian reference frame, allowing us to describe the system in a relational manner. More precisely, $\chi^0$ serves as a relational clock and $\chi^i$ with $i\in\{1,2,3\}$ serve as relational rods. The field $\phi$ will be assumed to dominate the energy-momentum budget of the system, being slightly inhomogeneous with respect to the rods.

\begin{figure}[h!]
    \centering
    \includegraphics[width=0.4\textwidth]{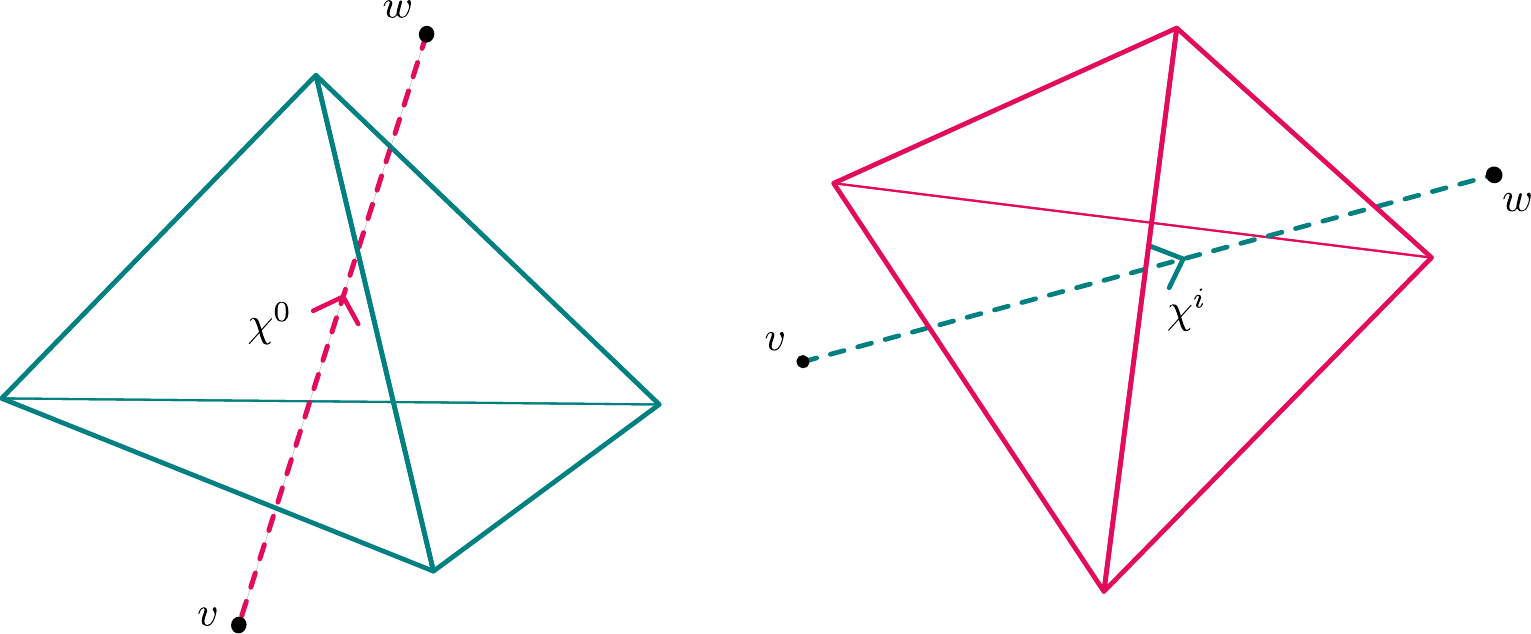}
    \caption{Left: clock $\chi^0$ propagating across spacelike tetrahedron (equivalently, along timelike dual edge) connecting the dual vertices $v$ and $w$. Right: rods $\chi^i$ propagating across timelike tetrahedron (equivalently, along spacelike dual edge) connecting the dual vertices $v$ and $w$. Propagation of matter data across spacelike (resp.\ timelike) tetrahedra, or timelike (resp.\ spacelike) dual edges  is encoded in the kinetic kernel $\mathcal{K}_+$ (resp.\ $\mathcal{K}_-$), hence the restrictions \eqref{eqn:restrictionkinetic}.}  
    \label{fig:spacetime_tetrahedra}
\end{figure}

Importantly, the manifestly causal nature of the minimally extended BC GFT model allows us, for the first time, to consistently implement the Lorentzian properties of the physical frame at the QG level. Indeed, at a classical, discrete geometric level, a clock propagates along timelike dual edges, while rods propagate along spacelike dual edges. These conditions can be imposed strongly at the quantum gravity level by requiring that (see Fig.~\ref{fig:spacetime_tetrahedra})
\begin{subequations}
\label{eqn:restrictionkinetic}
\begin{align}
\mathcal{K}_+&= \mathcal{K}_+\bigl(g_v,g_w; (\rf^0_v-\rf^{0}_w)^2, (\phi_v-\phi_w)^2\bigr),\label{eq:spacelike kernel restriction}\\[7pt]
\mathcal{K}_-&= \mathcal{K}_-\bigl(g_v,g_w;\abs{\vb*{\rf}_v-\vb*{\rf}_w}^2,(\phi_v-\phi_w)^2\bigr)\label{eq:timelike kernel restriction}.
\end{align}
\end{subequations}
Note that no such restriction is assumed for the matter field $\phi$.

\section{Entangled coherent peaked states}

Building on a series of previous results~\cite{Oriti:2016qtz,Marchetti:2020umh,Jercher:2021bie}, we suggest that the dynamics of scalar cosmological perturbations can be extracted from perturbed coherent states. The reasons for considering coherent states are twofold. First, they allow to systematically implement a mean-field approximation, and thus the simplest form of coarse-graining of the QG theory \cite{Oriti:2021oux}. Second, it has been shown \cite{Marchetti:2020umh,Pithis:2016cxg,Calcinari:2023sax} that coherent states lead to small relative fluctuations of geometric and matter field operators, which is essential for a semi-classical interpretation of the dynamics.  Furthermore, in virtue of their simple collective behavior, these states offer a transparent way of connecting macroscopic quantities to microscopic ones, as we will see explicitly below. More precisely, we consider states of the form
\begin{equation}\label{eqn:perturbedstates}
\ket{\Delta;x^0,\vb*{x}} = \mathcal{N}_\Delta e^{\hat{\sigma}\otimes\one_-+\one_+\otimes\hat{\tau}+\hat{\delta\Phi}\otimes\one_-+\hat{\delta\Psi}+\one_+\otimes\hat{\delta\Xi}}\ket{\emptyset},
\end{equation}
where $\mathcal{N}_\Delta$ is a normalization factor and $\ket{\emptyset} = \ket{\emptyset}_+\otimes\ket{\emptyset}_-$ is the $\mathcal{F}$-vacuum.
In the above expression, the one-body operators
\begin{equation}
\hat{\sigma}\equiv\sigma_{\vb*{\xi}_+;x^0,p_\phi}\cdot \hat{\varphi}^\dagger_+,\qquad 
\hat{\tau}\equiv \tau_{\vb*{\xi}_-;x^0,\vb*{x},p_\phi}\cdot\hat{\varphi}^\dagger_-,
\end{equation}
generate a background condensate state in the extended Fock space of the form
\begin{equation}
\ket{\sigma;x^0,p_\phi}\otimes\ket{\tau;x^0,\vb*{x},p_\phi} = \mathcal{N}_\sigma\mathcal{N}_\tau\e^{\hat{\sigma}\otimes\one+\one\otimes\hat{\tau}}\ket{\emptyset},
\end{equation}
with $\ket{\sigma;x^0,p_\phi}$ and $\ket{\tau;x^0,\vb*{x},p_\phi}$ representing spacelike and timelike condensates, respectively. The spacelike $\sigma_{\vb*{\xi}_+;x^0,p_\phi}$ and timelike $\tau_{\vb*{\xi}_-;x^0,\vb*{x},p_\phi}$ condensate wavefunctions are localized around $\chi^0=x^0$ and $\rf^0=x^0,\vb*{\rf}=\vb*{x}$, respectively. Furthermore, they are similarly localized in $\phi$-Fourier space ($\phi\rightarrow\pi_\phi$) around an arbitrary scalar field momentum $p_\phi$. In practice, this is achieved by assuming that they both factorize into fixed (Gaussian, see~\cite{Marchetti:2021gcv}) peaking functions and into reduced condensate wavefunctions, $\tilde{\sigma}$ and $\tilde{\tau}$, respectively.\footnote{We note that both the peaking properties of the states and the kinetic coupling \eqref{eqn:restrictionkinetic} manifestly reflect our choice of physical frame. A different choice of relational frame would affect both these aspects of our construction.} The peaking properties of the states are collectively represented by the multi-labels $\vb*{\xi}_\pm$, respectively, see~\cite{Jercher:2023nxa} for more details.

Requiring $\tilde{\sigma}$ (resp.\ $\tilde{\tau}$) to contain only gauge-invariant data, the spacelike (resp.\ timelike) condensate wavefunction can be seen as a distribution of geometric and matter data on a $3$--surface (resp.\ $(2+1)$--surface) localized at relational time $x^0$ (resp.\ relational point $(x^0,\vb*{x}))$. Therefore, averages of operators on such relationally localized states can be seen as effective relational observables~\cite{Marchetti:2020umh}. Relational homogeneity of the background structures is then imposed by assuming that $\tilde{\sigma}$ and $\tilde{\tau}$ only depend on the clock variable $\rf^0$. Furthermore, isotropy is imposed by requiring $\tilde{\sigma}$ and $\tilde{\tau}$ to depend only on a single spacelike\footnote{This is a non-trivial requirement for $\tilde{\tau}$, which can in principle carry spacelike ($\rho$) and timelike ($\nu$) representation labels~\cite{Jercher:2022mky,Dekhil:2024djp}.} representation label $\rho\in\mathbb{R}$. Under these assumptions, we have $\tilde{\sigma}=\tilde{\sigma}_\rho(\rf^0,\pi_\phi)$ and $\tilde{\tau}=\tilde{\tau}_\rho(\rf^0,\pi_\phi)$.

Inhomogeneities in \eqref{eqn:perturbedstates} are encoded in the operators $\hat{\delta\Phi}\otimes\one_-,\hat{\delta\Psi},$ and $\one_+\otimes\hat{\delta\Xi}$ which are in general $m$-body operators, with integer $m>1$. These produce quantum entanglement\footnote{This \qmarks{perspective-neutral} \cite{Vanrietvelde:2018pgb} entanglement is defined with respect to the kinematical (Fock) tensor product structure (TPS). This is expected to be inherited by the physical Hilbert space, due to the lack of redundant spacetime information in GFT. Thus, contrary to what happens for gravitational theories \cite{Giacomini:2017zju,Vanrietvelde:2018pgb,Hoehn:2019fsy,delaHamette:2020dyi,Hoehn:2023ehz}, one would expect this kinematical notion of entanglement to also be physical. In any case, we note that in this paper we do not go beyond the mean-field level. Thus, we work with the kinematical TPS, so that within our approximations, the above notion of entanglement is by construction physical.} within and between the spacelike and timelike sectors. For the remainder, we choose $m=2$, anticipating that, under our assumptions, this is sufficient to capture the physics of scalar, isotropic and slightly inhomogeneous cosmological perturbations. Higher $m$-body operators would offer more involved forms of entanglement and could be required for describing phenomenologically richer systems (e.g.\ involving anisotropies). However, following the arguments of~\cite{Jercher:2023nxa}, these would lead to further dynamical freedom discussed below and result in more involved dynamical equations which might be difficult to treat analytically. We therefore leave a treatment of such operators open for future research. Explicitly, the three $2$-body operators are defined as further
\begin{subequations}\label{eqn:entanglementkernels}
\begin{align}
    \hat{\delta\Phi}&=\hat{\varphi}^\dagger_+\cdot\delta\Phi\cdot\hat{\varphi}^\dagger_+,\\[4pt]
    \hat{\delta\Psi}&=\hat{\varphi}^\dagger_+\cdot\delta\Psi\cdot\hat{\varphi}^\dagger_-,\\[4pt] 
    \hat{\delta\Xi}&=\hat{\varphi}^\dagger_-\cdot\delta\Xi\cdot\hat{\varphi}^\dagger_-\,.
\end{align}  
\end{subequations}
The kernels $\delta\Phi$, $\delta\Psi$ and $\delta\Xi$ are in general bi-local, non-factorized (hence entangling) functions of the respective GFT field domains, determined by the mean-field dynamics we study below. Imposing isotropy\footnote{This condition, although restrictive, is compatible with the fact that we will only study isotropic geometric observables.} and gauge-invariance, the kernels take the form $\delta\Psi=\delta\Psi_{\rho_v\rho_w}(\rf^\mu_v,\rf^\mu_w,\phi_v,\phi_w)$, similarly for $\delta\Phi$ and $\delta\Xi$.

As we will see explicitly below, the above kernels, together with the reduced condensate wavefunctions, describe slightly inhomogeneous cosmological geometries. Their dynamics are derived via a mean-field approximation of the full quantum dynamics, i.e.
\begin{equation}\label{eq:lowest SDE}
\expval**{\fdv{S[\hat{\varphi}_\pm,\hat{\varphi_\pm}^{\dagger}]}{\hat{\varphi}(g_v,X_\pm,x^\mu,\mf)}}{\Delta;x^0,\vb*{x}}= 0.
\end{equation}
From here on we restrict ourselves to a mesoscopic regime of negligible GFT interactions~\cite{Oriti:2016qtz,Marchetti:2020umh} where only a single representation label $\rho_o$ dominates~\cite{Gielen:2016uft,Jercher:2021bie}. Thus, we will suppress any explicit dependence of functions on representation labels. Finally, since we are interested in small inhomogeneities represented by the kernels in \eqref{eqn:entanglementkernels}, we study the mean-field equations \eqref{eq:lowest SDE} perturbatively by working with linearized states
\begin{equation}\label{eq:linearized states}
\ket{\Delta;x^0,\vb*{x}} \approx \mathcal{N}_\Delta \left(\one+\hat{\delta\Phi}+\hat{\delta\Psi}+\hat{\delta\Xi}\right)e^{\hat{\sigma}\otimes\one_-+\one_+\otimes\hat{\tau}}\ket{\emptyset}.
\end{equation}
Neglecting interactions and simultaneously working perturbatively collapses the whole set of Schwinger-Dyson equations to the lowest order equations~\eqref{eq:lowest SDE}, see~\cite{Jercher:2023nxa}. As we will discuss below, this will result in a dynamical freedom at the level of perturbations. 

We now proceed to the perturbative study of equations \eqref{eq:lowest SDE}.
At the background level, and in the limit of negligible interactions, equations for the spacelike and timelike sectors exactly decouple. Following from the peaking in $\pi_\phi$ and in the frame variables, the equations of motion of the spacelike $\slrcw$ and timelike $\tlrcw$ reduced condensate wavefunctions become second order differential equations\footnote{Coefficients of different derivative terms are obtained from a power expansion of the kernels $\mathcal{K}_{\pm}$.} in relational time. In the limit in which the modulus of these reduced condensate wavefunctions is large (associated with late relational times and classical behavior, see \cite{Oriti:2016qtz,Pithis:2016cxg,Marchetti:2020qsq}), solutions to the equations of motion are given by
\begin{subequations}\label{eq:bkgcondensates}
\begin{align}
\slrcw(x^0,\pmm) &= \slrcw_0\e^{(\mu_+ + i\tpip)x^0},\label{eqn:bkgsigma}\\
\tlrcw(x^0,\pmm) &= \tlrcw_0\e^{(\mu_- + i\tpim)x^0},\label{eqn:bkgtau}
\end{align}
\end{subequations}
where $\mu_\pm=\mu_\pm(p_\phi)$ are in general functions of the peaking value of the matter field momentum.

At first order in perturbations, we obtain two differential equations for the three kernels $\delta\Psi$, $\delta\Phi$ and $\delta \Xi$. This dynamical freedom cannot be reduced beyond mean-field as long as interactions are negligible and the above kernels are small. However, it can be completely fixed by requiring a low-energy agreement with classical physics, as will be seen below. We make an ansatz
\begin{equation}\label{eq:relation of dPsi and dPhi}
\delta\Phi(\rf^\mu,\mm) = \mathrm{f}(\rf^\mu)\delta\Psi(\rf^\mu,\mm),  
\end{equation}
with the complex function $\mathrm{f}$ given by
\begin{equation}\label{eq:def of f}
\mathrm{f}(\rf^0,\vb*{\rf}) = f(\rf^0)\e^{i\theta_f(\rf^0)}\abs{\eta_\delta(\abs{\vb*{\rf}-\vb*{x}})}\e^{2i\pip\rf^0},
\end{equation}
where $f$ and $\theta_f$ are arbitrary real functions of $\rf^0$ and $\eta_\delta$ is a Gaussian peaking function with width $\delta$. Moreover, from now on, we consider correlations that belong to the same relationally localized $4$-simplex and are $\pi_\phi$-conserving, i.e.
\begin{equation}\label{eq:locality condition}
\delta\Psi = \delta\Psi(\rf^\mu,\mm)\delta^{(4)}(\rf^\mu-\rf^{\prime\mu})\delta(\mm-\mm').
\end{equation}
Assuming specific relations of the peaking parameters, see~\cite{Jercher:2023nxa}, the perturbation equations can be simplified further. In particular, the kernel $\delta\Psi$ entangling timelike and spacelike sectors, satisfies an equation of the form
\begin{equation}
0 = \delta\Psi''+t_1\delta\Psi'+t_0\delta\Psi+s_2\nabla_{\vb*{x}}^2\delta\Psi,
\end{equation}
where primes denote relational time derivatives and $t_i\equiv t_i[f,\theta_f](x^0,\pmm)$ and $s_2[f,\theta_f](x^0,\pmm)$ are complex quantities depending on the functions $f$ and $\theta_f$. This form of the equations already bears resemblance with the perturbations equations of GR, the explicit connection of which we develop in the next section.

\section{Effective dynamics of cosmological scalar perturbations}\label{sec:dynamics}

As mentioned above, at late times, quantum fluctuations of extensive operators acting on the Fock space $\mathcal{F}$ are small~\cite{Marchetti:2020qsq}. In this regime, which we consider from here on, one can
associate classical cosmological quantities $\mathcal{O}$ with expectation values of appropriate one-body GFT operators $\hat{\mathcal{O}}$ on the states \eqref{eq:linearized states}, defined as
\begin{align}\label{eqn:observablesgen}
\mathcal{O}(x^0,\vb*{x})&\equiv \expval{\hat{\mathcal{O}}}{\Delta;x^0,\vb*{x}}\nonumber\\
&= \bar{\mathcal{O}}(x^0)+\delta\mathcal{O}(x^0,\vb*{x}).
\end{align}
The above expectation value is effectively localized in relational spacetime, and thus should be compared to a corresponding classical relational \cite{Goeller:2022rsx} (or, equivalently, harmonic gauge-fixed) observable \cite{Gielen:2018fqv}. By construction, \eqref{eqn:observablesgen} splits into a background, $\bar{\mathcal{O}}$, and a perturbation, $\delta\mathcal{O}$. Below we compute equation \eqref{eqn:observablesgen} for some important geometric and matter observables, and we study their dynamics in detail.

A crucial example of such observables is the spatial $3$-volume $\hat{V}$ \cite{Oriti:2016qtz}, the expectation value $V_\Delta$ of which splits into a background contribution $\bar{V}(x^0,\pmm)= v\abs{\slrcw(x^0,\pmm)}^2$ and a perturbation
\begin{equation}
\delta V(x^0,\pmm)=2v\Re\{F[\delta\Psi](x^\mu,\pmm)\}]\,,
\end{equation}
where $v$ is a volume eigenvalue~\cite{Jercher:2021bie} scaling as $v\sim \rho_o^{3/2}$ and $F$ is a functional depending on background wavefunctions as well as the functions $f$ and $\theta_f$. The background volume satisfies
\begin{equation}
\frac{\bar{V}'}{3\bar{V}} = \frac{2}{3}\mu_+(\pmm),\qquad \left(\frac{\bar{V}'}{3\bar{V}}\right)' = 0,
\end{equation}
which successfully matches classical flat Friedmann dynamics in harmonic gauge if $\mu_+(\pmm)=3\bar{\pi}_\phi/(8M_{\Pl}^2)$~\cite{Oriti:2016qtz,deCesare:2016axk,Marchetti:2021gcv}. Since the spacelike and timelike sector decouple at background level, contributions of timelike tetrahedra drop out for spacelike observables.

A similar matching can be performed for the perturbations, which completely fixes the functions $f$ and $\theta_f$ (see~\cite{Jercher:2023nxa} for further details). In this way, the perturbed volume dynamics take the simple form
\begin{equation}\label{eqn:pertvolume}
\left(\frac{\delta V}{\bar{V}}\right)'' +a^4 k^2\left(\frac{\delta V}{\bar{V}}\right) = -3\mathcal{H}\left(\frac{\delta V}{\bar{V}}\right)',
\end{equation}
where $\mathcal{H}=\bar{V}'/(3\bar{V})$ is the Hubble parameter and $k$ is the Fourier mode relative to $\vb*{x}$. The harmonic term entering with $a^4$ constitutes an essential improvement compared to previous work~\cite{Marchetti:2021gcv} and is a combined consequence of the Lorentzian reference frame and the use of entangled CPSs. The right-hand side of Eq.~\eqref{eqn:pertvolume} is reminiscent of a friction term, which may be associated with a macroscopic dissipation phenomenon into the quantum gravitational microstructure (as suggested e.g.\ in~\cite{Perez:2017krv}).

Note that as a consequence of the matching, $\delta\Xi$ is only time-dependent. This in turn implies that one can reabsorb perturbations in the number of timelike quanta in the background component.\footnote{We note however, that timelike perturbations may indeed be relevant if one were to impose classicality conditions on observables that are not purely spacelike.}

Analogously, one can study expectation values of the scalar field operators $\hat{\Upphi}_\pm$ (and their conjugate momenta $\hat{\Pi}_\pm$), defined on each of the two sectors~\cite{Jercher:2023nxa}, to identify a matter scalar field $\mf_\Delta$. Since classically the matter field is an intensive quantity, we combine the expectation values $\Upphi_\pm$ of the scalar field operators $\hat{\Upphi}_\pm$ through the following weighted sum\footnote{This is analogous to how intensive quantities such as chemical potentials are combined in statistical physics, see \cite{Callen}.}
\begin{equation}\label{eq:mf as weighted sum}
\mf_\Delta = \Upphi_+\frac{N_+}{N}+\Upphi_-\frac{N_-}{N},
\end{equation}
where $N$ is the total (average) number of quanta, and $N_\pm$ are the (average) number of quanta in each sector. The above quantity can then be split in a background, $\bar{\mf}$, and perturbed component, $\delta\mf$.

At the background level, one can show that by requiring $\bar{\Upphi}_\pm$ to be intensive quantities and assuming $\mu_+>\mu_-$ (which enter the background condensate solutions in Eqs.~\eqref{eq:bkgcondensates}), $\bar{\mf}=\bar{\Upphi}_+$ is completely captured by spacelike data at late times and satisfies the classical equation of motion $\bar{\mf}''=0$. Moreover, the background matter analysis unambiguously identifies the peaking momentum value $p_\phi$ with the classical background momentum of the scalar field, $\bar{\pi}_\mf$ \cite{Marchetti:2021gcv}. 

At first order in perturbations, and under the same assumptions as above, one can write
\begin{equation}\label{eqn:defpertmf}
\delta\phi=\left(\frac{\delta N_+}{\bar{N}_+}\right)\bar{\phi}=\left(\frac{\delta V}{\bar{V}}\right)\bar{\mf},
\end{equation}
so that, using \eqref{eqn:pertvolume} and $\bar{\phi}''=0$, we obtain
\begin{equation}\label{eqn:pertmf}
\delta\mf''+a^4 k^2\delta\mf = J_{\mf},
\end{equation}
with the source term $J_{\mf}$ given in Eq.~\eqref{eq:phi sourceterm}. 

A crucial quantity in classical cosmology is the comoving curvature perturbation $\mathcal{R}$~\cite{lyth}, proportional to the so-called Mukhanov-Sasaki variable~\cite{Mukhanov:1988jd,mukhanov2,sasaki}. This can be obtained by combining matter and geometric (non-exclusively volume) information in a gauge-invariant way. Restricting the geometric data to volume only, one can define an analogous \qmarks{curvature-like} variable
\begin{equation}
\tilde{\mathcal{R}}\equiv  -\frac{\delta V}{3\bar{V}}+\mathcal{H}\frac{\delta\mf}{\bar{\mf}'},
\end{equation}
which is perturbatively gauge-invariant only in the super-horizon limit. The \qmarks{curvature-like} variable $\tilde{\mathcal{R}}$ can be constructed within our framework by combining equations \eqref{eqn:pertvolume} and \eqref{eqn:pertmf}\footnote{We note that in this case $\tilde{\mathcal{R}}$ is constructed out of (effectively) relational observables, and thus is gauge invariant by construction.} and satisfies
\begin{equation}\label{eq:GFT perturbed R}
\tilde{\mathcal{R}}'' + a^4 k^2\tilde{\mathcal{R}} = J_{\tilde{\mathcal{R}}},
\end{equation}
where $J_{\tilde{\mathcal{R}}}$ is presented in Eq.~\eqref{eq:R sourceterm}. 

Alternatively, Eq.~\eqref{eq:GFT perturbed R} can be recast in conformal time, commonly used in standard cosmology, by introducing a harmonic parametrization of the reference fields and changing to conformal time $\tau$ via $\dd{\tau} = a^2\dd{x^0}$, yielding
\begin{equation}
\dv[2]{\tilde{\mathcal{R}}}{\tau}+2H\dv{\tilde{\mathcal{R}}}{\tau}+k^2\tilde{\mathcal{R}} = J_{\tilde{\mathcal{R}}}(\tau,k),
\end{equation}
where $H$ is the conformal Hubble parameter and $J_{\tilde{\mathcal{R}}}(\tau,k)$ is given in Eq.~\eqref{eq:R sourceterm CL}.

\begin{figure}[h!]
    \centering
    \includegraphics[width=0.45\textwidth]{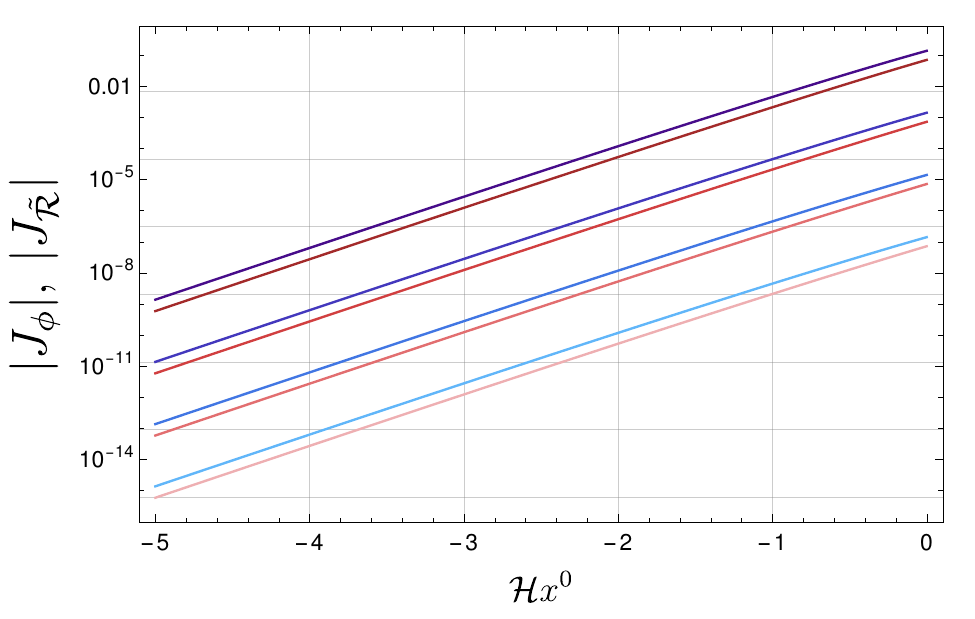}
    \caption{Absolute value of source terms $J_\mf$ (blue) and $J_{\tilde{\mathcal{R}}}$ (red) with $c = 0.1$ for modes $k/M_{\Pl}\in\{10^{-3},\: 10^{-2},\: 10^{-1},\: 10^0\}$ in the vicinity of the trans-Planckian regime where darker colors correspond to larger modes. In the trans-Planckian regime $k/M_{\Pl} \sim 1$, $J_\mf$ and $J_{\tilde{\mathcal{R}}}$ become of order one at late times, while they remain small at all times in the sub-Planckian regime.}
    \label{fig:sourceterms}
\end{figure}

Comparing equations \eqref{eqn:pertmf} and \eqref{eq:GFT perturbed R} with their classical GR counterparts \eqref{eq:clperturbedmf} and~\eqref{eq:clperturbedR}, we notice that they in general contain an additional source term, $J_\mf$ and $J_{\tilde{\mathcal{R}}}$, respectively. The intrinsically quantum gravitational nature of these terms can be made manifest by solving first Eq.~\eqref{eqn:pertvolume} under the requirement that solutions match the GR ones in the super-horizon limit. Indeed, in this case we can write
\begin{equation}\label{eqn:sourceterms}
    J_{\mf}=c\left(\frac{a^2k}{M_{\Pl}}\right)j_{\mf}[\bar{\mf}]\,,\qquad J_{\tilde{\mathcal{R}}}=c\left(\frac{a^2k}{M_{\Pl}}\right)j_{\tilde{\mathcal{R}}}[\bar{\mf}]\,,
\end{equation}
where $c$ is the initial ratio of perturbed and background volume (and thus required to be small) and $j_{\mf}$ and $j_{\tilde{\mathcal{R}}}$ are functions depending on the background field $\bar{\mf}$ and the mode $k$. Visualized in Fig.~\ref{fig:sourceterms}, effects of the source terms remain small at all times in the sub-Planckian regime $k/M_{\Pl}\ll 1$ such that classical dynamics are recovered. Effects of the source terms become important in the trans-Planckian regime, $k/M_{\Pl}\gtrsim 1$. Thereby, the results of this letter go well beyond~\cite{Jercher:2023nxa} in that they prove the corrections to be genuine quantum gravity effects.

\begin{figure}[h!]
    \centering
    \includegraphics[width=0.45\textwidth, trim = 20pt 200pt 100pt 100pt, clip]{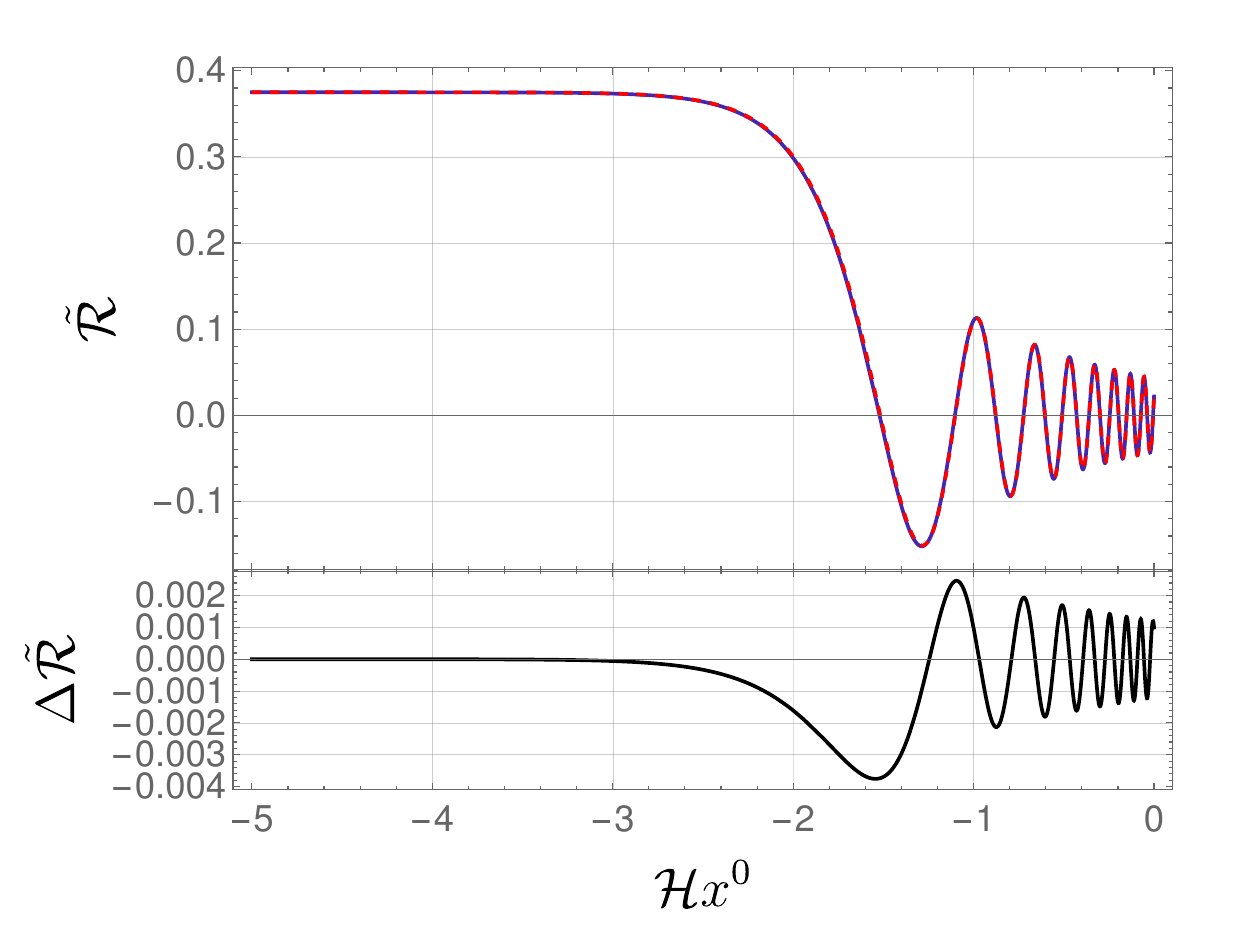}
    \caption{Top: Curvature-like perturbation $\tilde{\mathcal{R}}_\GFT$ (blue) and $\tilde{\mathcal{R}}_\GR$ (dashed red) for a fixed mode $k/M_\Pl = 10^2$ in the trans-Planckian regime. Bottom: Difference of the blue and red dashed curves above, $\Delta\tilde{\mathcal{R}}\equiv \tilde{\mathcal{R}}_\GFT-\tilde{\mathcal{R}}_\GR$. For both plots, the ratio of $\delta V/\bar{V}$ and $\delta\mf$ at initial time has been set to $c=0.1$.}
    \label{fig:Rplot}
\end{figure}

An exemplary solution of Eq.~\eqref{eq:GFT perturbed R}, together with a comparison to the classical GR counterpart is depicted in Fig.~\ref{fig:Rplot}. As it is clear from the plots, QG corrections are only relevant for highly trans-Planckian modes, which are many orders of magnitude larger than the typical momentum scales of interest in cosmology ($\sim \text{Mpc}^{-1}$). 
It would be particularly interesting to investigate whether the above corrections can also be obtained from a continuum modified gravitational action, potentially establishing an effective field theory framework for the above quantum dynamics. We leave this intriguing avenue for future work.

\section{Conclusions}

In this letter, we provided a concrete example of the emergence of geometry from quantum gravity entanglement~\cite{Giddings:2005id,Ryu:2006bv,VanRaamsdonk:2010pw,Bianchi:2012ev,Maldacena:2013xja,Giddings:2015lla,Cao:2016mst,Donnelly:2016auv,swingle2018spacetime,colafrancheschi2022emergent,Bianchi:2023avf}, 
by describing cosmological inhomogeneities in terms of relational nearest-neighbor two-body correlations between GFT quanta of geometry, encoded in entangled GFT coherent states. Our results therefore offer crucial insights into the potential intrinsically quantum gravitational nature of cosmological perturbations. More precisely, we have shown that these correlations produce macroscopic inhomogeneities in the expectation values of collective observables on the above states, such as the perturbed volume, $\delta V$, matter, $\delta\mf$, and the curvature-like perturbation $\tilde{\mathcal{R}}$. From this perspective, the question of the origin of the slight inhomogeneity of the universe could be addressed by understanding why low entanglement states are dynamically preferred by the underlying underlying QG theory. As the main result of this work, we extracted the dynamics of these cosmological perturbations from full QG (as described by the causally complete BC model), and showed that there are modifications with respect to the classical results. Crucially, these modifications, captured by the source terms $J_\mf$ and $J_{\tilde{\mathcal{R}}}$, are negligible on sub-Planckian scales and show relevance only in the trans-Planckian regime. Consequently, the modifications constitute genuine quantum effects arising from the microscopic spacetime structure. 

There are two different levels at which one could try to make contact with observations. First, one could phenomenologically incorporate the GFT modified perturbation dynamics into the standard cosmological model. Since inflation can be sensitive to trans-Planckian physics~\cite{martin2001, Danielsson:2002kx, Brandenberger:2012aj}, this may already produce observable consequences. On this note, it is worth pointing out that the GFT-induced trans-Planckian corrections to the perturbation dynamics appear as source terms rather than modified dispersion relations (which is the scenario that has been mostly investigated in the literature \cite{Brandenberger:2000wr, Brandenberger:2002hs, Starobinsky:2001kn, Niemeyer:2001qe, Kempf:2001fa}). While inflation has been shown to be robust against the latter corrections, as long as scales are sufficiently separated (i.e. $H/M_{\text{Pl}}\ll 1$ during the inflationary phase)~\cite{Niemeyer:2001qe}, to the best of our knowledge, there is no analogous argument for the former.

Alternatively, one could \emph{derive} the physics of cosmological perturbations (including the properties of their power spectra) from the full QG theory. However, this would require: (i) the construction of additional \emph{relational} operators capturing non-isotropic and non-scalar perturbations, (ii) the inclusion of a more realistic matter content, (iii) the generalization of the analysis performed here to early times close to the quantum bounce~\cite{Oriti:2016qtz,Oriti:2016ueo,Marchetti:2020umh,Jercher:2021bie}, (iv) the inclusion of an inflationary mechanism (which may also very well emerge from the QG theory), and (v) some degree of control over the quantum field theory limit of the underlying QG theory. For an extensive discussion of these points, we refer the reader to~\cite{Jercher:2023nxa}.

It could also be interesting to further study and differentiate possible types of entanglement at play in our setting. For instance, within the context of the previous paragraph one could try to extract the entanglement of the scalar perturbations and compare them with computations in the continuum~\cite{Brahma:2020zpk}. Moreover, as recent works suggest~\cite{Bose:2017nin,Marletto:2017kzi,Mehdi:2022oct,Christodoulou:2022mkf}, gravity, should it really be quantized, mediates entanglement of matter degrees of freedom. These ideas were then used to show that the curvature of an expanding background should leave fingerprints on the corresponding entanglement profile~\cite{Brahma:2023lqm}. We leave it to future work to study if and how entanglement of (continuum) matter degrees of freedom could be mediated by the quantum gravity model used here.

While we realized this program in the context of the BC GFT model, given recent results suggesting that the BC and EPRL~\cite{Rovelli:2011eq,Perez:2012wv,Conrady:2010kc,Conrady:2010vx} models could lie in the same universality class~\cite{Asante:2020qpa,Jercher:2021bie,Dittrich:2021kzs,Dittrich:2022yoo,Marchetti:2022igl,Marchetti:2022nrf,Dekhil:2024ssa,Dekhil:2024djp}, we expect our results to also transfer to the latter model. Moreover, we anticipate that our findings will impact on a wider set of quantum gravity approaches that incorporate a causally complete set of discrete Lorentzian geometries. Distinguishing different such quantum gravity models by their phenomenological implications is an intriguing question that we leave to future research. A promising approach could be to include tensor perturbations in the present framework to further constrain the dynamics, potentially yielding quantitatively or even qualitatively different cosmological predictions, see also Ref.~\cite{Jercher:2023nxa} for a discussion of this matter.

Finally, note that the extended causal structure of the complete BC model plays a crucial role in the derivation of the above results. Indeed, it is not only important for the construction of the entangled CPS, but it also allows for a consistent coupling of the physical Lorentzian reference frame, whose causal properties are a necessary ingredient to recover the classical dynamics in a low-energy limit. Seen in a broader context, these results strengthen the arguments for models that incorporate a causally complete set of discrete geometries such as causal dynamical triangulations~\cite{Ambjorn:2012jv,Loll:2019rdj,Loll:1998aj}, causal sets~\cite{Surya:2019ndm}, Lorentzian Regge calculus~\cite{Jercher:2023csk,Asante:2021phx,Asante:2021zzh,Dittrich:2021gww,Dittrich:2023rcr} or Lorentzian spin foams and LQG~\cite{Conrady:2010kc,Conrady:2010vx,Simao:2021qno,Han:2021bln,Han:2021kll}. Given the strong relations between GFT and this broader class of quantum gravity approaches based on discrete structures, we hope that our results serve as an encouragement to employ a physical Lorentzian reference frame and to explore the relation between quantum geometric entanglement and cosmological perturbations in the latter.

\begin{acknowledgments}
\textbf{Acknowledgements} The authors would like to thank two anonymous referees and Jos\'{e} Sim\~{a}o for their insightful and constructive comments on the manuscript. The authors are grateful to Edward Wilson-Ewing, Daniele Oriti, Kristina Giesel, Steffen Gielen, Renata Ferrero, Viqar Husain and Mairi Sakellariadou for helpful discussions and comments. AFJ gratefully acknowledges support by the Deutsche Forschungsgemeinschaft (DFG, German Research Foundation) project number 422809950 and by grant number 406116891 within the Research Training Group RTG 2522/1. AGAP acknowledges funding from the DFG research grants OR432/3-1 and OR432/4-1 and the John-Templeton Foundation via research grant 62421. AFJ and AGAP are grateful for the generous financial support by the MCQST via the seed funding Aost 862933-9 granted by the DFG under Germany’s Excellence Strategy – EXC-2111 – 390814868. AGAP gratefully acknowledges support by the Center for Advanced Studies (CAS) of the LMU through its \emph{Junior Researcher in Residence program}. LM acknowledges support from AARMS and the Natural Sciences and Engineering Research Council of Canada, and thanks the Arnold Sommerfeld Center (LMU) for hospitality.
\end{acknowledgments}

\appendix
\section{Appendix}
\subsection{Classical perturbation equations in harmonic gauge}

The line element of scalar perturbations around a flat FLRW background in harmonic coordinates is given by
\begin{widetext}
\begin{equation}
\dd{s}^2 = -a^6(1+2A)\dd{t}^2+a^4\partial_iB\dd{t}\dd{x^i}+a^2\left((1-2\psi)\delta_{ij}+2\partial_i\partial_j E\right)\dd{x}^i\dd{x}^j.
\end{equation}
\end{widetext}

Via the local spatial volume element, we identify the background and perturbed volume as
\begin{equation}\label{eq:clvolumeexpr}
\bar{V} = a^3,\qquad \frac{\delta V}{\bar{V}} = -3\psi+\nabla^2 E.
\end{equation}
Together with a minimally coupled massless scalar field $\mf=\bar{\mf}+\delta\mf$ with conjugate momentum $\pi_{\phi}$, background and perturbed equations of motion of the volume and the scalar field are given by
\begin{equation}
3 \left(\frac{\bar{V}'}{3\bar{V}}\right)^2 = \frac{1}{2 M_{\Pl}^2}\bar{\pi}_\mf^2,
\end{equation}
and
\begin{align}
\left(\frac{\delta V}{\bar{V}}\right)''-a^4 \nabla^2\left(\frac{\delta V}{\bar{V}}\right) &= 0,\label{eq:pert relative volume}\\[7pt]
\delta\mf''-a^4\nabla^2\delta\mf &= 0,\label{eq:clperturbedmf}
\end{align}
respectively.

To compare classical dynamics with that of GFT, we define the \textit{curvature-like perturbation} $\tilde{\mathcal{R}}$
\begin{equation}\label{eq:classical Rtilde}
\tilde{\mathcal{R}}\equiv -\frac{\delta V}{3\bar{V}}+\mathcal{H}\frac{\delta\mf}{\bar{\mf}'},
\end{equation}
obeying
\begin{equation}\label{eq:clperturbedR}
\tilde{\mathcal{R}}''-a^4\nabla^2\tilde{\mathcal{R}} = 0.
\end{equation}
Notice that in classical GR, $\tilde{\mathcal{R}}$ is explicitly gauge-dependent.

\subsection{Source terms}

Depicted in Fig.~\ref{fig:sourceterms}, the source terms entering the perturbed matter and curvature-like equations,~\eqref{eqn:pertmf} and~\eqref{eq:GFT perturbed R}, respectively, are given by
\begin{subequations}
\begin{align}
 J_{\mf} &= \left(-3\mathcal{H}\bar{\phi}+2\bar{\phi}'\right)\left(\frac{\delta V}{\bar{V}}\right)',\\[7pt]
 J_{\tilde{\mathcal{R}}} &= \left[3\mathcal{H}-\frac{1}{4 M_{\Pl}^2}\left(\bar{\mf}^2\right)'\right]\left(\frac{\delta V}{\bar{V}}\right)'.
\end{align}
\end{subequations}
For solutions of $\delta V/\bar{V}$ that match GR in the super-horizon limit, these terms explicitly evaluate to
\begin{subequations}\label{eq:explicitsourceterms}
\begin{align}
    & J_\mf = -c\left(\frac{a^2 k}{M_{\Pl}}\right)\bar{\mf}'\left[-\frac{3}{\sqrt{6}}\bar{\mf}+2 M_{\Pl}\right]\Gamma\left(\frac{7}{4}\right)\nonumber\\[4pt]
    &\quad\:\times\left(\frac{4\mathcal{H}}{a^2 k}\right)^{3/4}J_{7/4}\left(\frac{a^2 k}{2\mathcal{H}}\right)\,,\label{eq:phi sourceterm}\\[7pt]
    & J_{\tilde{\mathcal{R}}} = -c\left(\frac{a^2 k}{M_{\Pl}}\right)\bar{\mf}'\left[\frac{3}{\sqrt{6}}-\frac{\bar{\mf}}{2M_{\Pl}}\right] \Gamma\left(\frac{7}{4}\right)\nonumber\\[4pt]
    &\quad\:\times\left(\frac{4\mathcal{H}}{a^2 k}\right)^{3/4}J_{7/4}\left(\frac{a^2 k}{2\mathcal{H}}\right),\label{eq:R sourceterm}
\end{align}
\end{subequations}
where $c$ is an initial condition and $J_{\alpha}$ are Bessel functions of the first kind.

Alternatively, the source terms can be expressed in terms of conformal-longitudinal coordinates. In particular, $J_{\tilde{\mathcal{R}}}(\tau,k)$ is given by
\begin{equation}
J_{\tilde{\mathcal{R}}}(\tau,k) = \left[3H-\frac{1}{4 M_{\Pl}^2}\dv{}{\tau}\left(\bar{\mf}^2\right)\right]\dv{}{\tau}\left(\frac{\delta V}{\Bar{V}}\right),
\end{equation}
which, upon solutions of $\delta V/\bar{V}$, is defined as
\begin{align}\label{eq:R sourceterm CL}
J_{\tilde{\mathcal{R}}}(\tau,k) &= -c\left(\frac{k}{M_{\Pl}}\right)\dv{\bar{\mf}}{\tau}\left[\frac{3}{\sqrt{6}}-\frac{\bar{\mf}}{2 M_{\Pl}}\right]\Gamma\left(\frac{7}{4}\right)\nonumber\\[4pt]
&\times\left(\frac{2}{k\tau}\right)^{3/4}J_{7/4}(k\tau).
\end{align}   

\bibliography{references.bib} 

\end{document}